\documentclass[%
 reprint,
 amsmath,amssymb,
 aps,
prb,
]{revtex4-2}

\usepackage{graphicx}
\usepackage{dcolumn}
\usepackage{bm}

\usepackage[dvipsnames]{xcolor,colortbl}

\begin{document}

\preprint{APS/123-QED}

\title{Phase separation dynamics in a symmetric binary mixture of ultrasoft particles}

\author{Tanmay Biswas}
\affiliation{Institut f\"ur Theoretische Physik, TU Wien, Wiedner Hauptstrasse 8-10, A-1040 Wien, Austria}

\author{Gerhard Kahl}%
 \email{gerhard.kahl@tuwien.ac.at}
\affiliation{Institut f\"ur Theoretische Physik, TU Wien, Wiedner Hauptstrasse 8-10, A-1040 Wien, Austria}

\author{Gaurav P. Shrivastav}
\affiliation{Institut f\"ur Theoretische Physik, TU Wien, Wiedner Hauptstrasse 8-10, A-1040 Wien, Austria}

\date{\today}

\begin{abstract}

Phase separation plays an  role in determining the self-assembly of biological and soft-matter systems. In biological systems, liquid-liquid phase separation inside a cell leads to the formation of various macromolecular aggregates. The interaction among these aggregates is soft, i.e., these can significantly overlap at a small energy cost. From the computer simulation point of view, these complex macromolecular aggregates are generally modeled by the so-called soft particles. The effective interaction between two particles is defined via the generalized exponential potential (GEM-$n$) with $n = 4$. Here, using molecular dynamics simulations, we study the phase separation dynamics of a size-symmetric binary mixture of ultrasoft particles. We find that when the mixture is quenched to a lower temperature below the critical temperature, the two components spontaneously start to separate. Domains of the two components form, and the equal-time order parameter reveals that the domains grow in a power-law manner with exponent $1/3$, which is consistent with the Lifshitz-Slyozov law for conserved systems. Further, the static structure factor shows a power-law decay with exponent $4$ consistent with the Porod law.

\end{abstract}

\maketitle



\section{\label{sec:intro}Introduction}

In many biological and polymeric systems liquid-liquid phase separation in binary (or more component) systems plays an eminent role in triggering self-assembly processes and the morphology of systems (see, e.g., \cite{hyman2014liquidphaseseparation, bates1991polymerphase}). In biological cells, such phase separation scenarios are induced by the differences in the molecular affinity or chemical potential of the species, thus macromolecules condense into high- and low-density phases (see, for example, \cite{banani2017biomolecularcondens, uversky2017disorderedproteins}). These biological systems are in general characterized by a high density and the effective interaction among the constituent molecular assemblies can be considered as soft, allowing thereby significant spatial overlap between the entities \cite{Klapp_2004,Denton2007}.

In computer simulations, such mutually penetrable, complex macromolecular aggregates can conveniently be modeled via so-called ultrasoft particles, i.e., potentials that are characterized by a finite energy penalty at zero separation. These effective interactions emerge as the degrees-of-freedom of the constituting microscopic particles (such as atomic or molecular entities) are traced out \cite{Klapp_2004, Denton2007}. Among these ultrasoft interactions the so-called generalized exponential model (with index $n$) \cite{mladek2006formation}-- GEM-$n$ (with typically $n \ge 2$) -- ranges among the most familiar ones, due to the simplicity of its functional form and its wide-spread use in various investigations (see, e.g., \cite{mladek2006formation,mladek2007clustering,likos2008cluster,mladek2008multiple,mladek2007phase,likos2007ultrasoft,zhang2010reentrant,gaurav2020,gaurav2021softmatter}. {
For $n > 2$ GEM-$n$ particles form -- despite their mutual repulsion -- stable clusters of overlapping particles: these clusters are polydisperse in the disordered phase and condense upon increasing the densities in stable cluster crystals, i.e., BCC or FCC lattices, where each lattice site is populated by a well-formed cluster of overlapping particles. The phase diagram of such systems (notably for $n= 4$) is well-explored both at high \cite{mladek2006formation, mladek2007clustering,
mladek2007clustering,likos2008cluster} as well as at low temperatures \cite{zhang2010reentrant}. The mechanism behind this unusual phase behaviour and of various other interesting properties \cite{mladek2007phase} is  well-understood and documented \cite{likos2007ultrasoft}.
The particle dynamics in the crystalline state is even more intriguing as particles keep hopping from one cluster to another, preserving the overall crystalline structure \cite{coslovich2011hopping}. Thus cluster crystals represent an interesting model for defect-rich clusters with interesting structural and mechanical properties (see, e.g., \cite{gaurav2021softmatter,gaurav2020,ganguly2022}.

Ever since the publication of Ref. \cite{mladek2006formation} a few attempts have been made to extend investigations of such a system to more component mixtures of ultrasoft particles. Molecular dynamics simulations have revealed that a polydisperse mixture of ultrasoft particles shows a {\it cluster glass phase}: the structure of the clusters (represented via the positions of their centers of mass of clusters) remains disordered and their dynamics becomes drastically slowed down, both features that represent signatures of the emergence of a glassy phase \cite{coslovich2012clusterglass}. On the other hand, a size-asymmetric binary mixture of ultrasoft particles shows micro-segregation of the two components in the clusters \cite{coslovich2012clusterglass}. Finally, a few investigations on the ordered phases of binary mixture can be reported (see, e.g., , despite clear evidence of the emergence of cluster crystals for such systems \cite{tscharnutter2024,scacchi2021}.

However, phase separation in these mixtures has not yet been studied, a phenomenon which is of utmost importance in understanding the self-assembly occuring in many biological and polymeric systems induced by phase separation. Also, the {\it kinetics of phase separation}, i.e., the evolution of the dynamics of the system when rendered thermodynamically unstable by a sudden change of temperature, has not been studied, yet. Several theoretical, simulation, and experimental studies have been dedicated to understand phase separation kinetics in polymer blends and soft-colloid mixtures (see, e.g., \cite{gennes1980dynamics, binder1983collective, kwak1992electron}). However none of these have dealt with the phase separation kinetics of a mixtures of ultrasoft particles for which interesting phenomena can be expected due to their cluster-forming capacity.

In this contribution we study the phase separation dynamics of a size-symmetric, equimolar binary mixture of ultrasoft particles (with species A and B). Using molecular dynamics simulations \cite{plimpton1995fast} in the canonical ensemble and considering a fairly large number of particles, we determine the tentative phase diagram of this mixture in the temperature-density plane and identify the coexistence curve and an estimate for the critical temperature $T_{\rm c}$. At high temperatures, the two components of the mixture remain in a homogeneously mixed state, while below $T_{\rm c}$ the species spatially separate into A-rich and B-rich phases. The two phases are identified by recording the spatial distribution of the difference in the concentrations of A and B particles  \cite{subir2006static}. When quenched from a high-temperature mixed state to a subcritical temperature, the mixture spontaneously phase separates, and domains of A and B-rich regions are formed. In an effort to quantify this dynamic process we correlate at the same instant the local concentrations at two locations in the system which are separated by a distance $r$ via a correlation function $C(r, t)$. This function provides evidence of a self-similar growth (when scaled with a time-dependent length $l(t)$ which estimates the size of the emerging clusters) which shows asymptotically for large distances $r$ a power-law growth with an exponent equal to $1/3$ \cite{puribook}. This feature is consistent with the predictions of the Lifshitz–Slyozov law for conserved systems \cite{LS1961kinetics} and has been extensively studied for Lennard-Jones fluids \cite{shaista2007crossover, md2017, md2011, bsg2022}. Furthermore, the spatial Fourier transform of this correlation function, a time-dependent structure factor shows at high wave vectors a power-law decay with an exponent equal to $4$, a feature which is consistent with Porod's law which signifies the presence of sharp interfaces in the system \cite{porod1982, puri1988large, gps2014}. 

The remainder of the manuscript is organized as follows: in Section~\ref{sec:model_simulations}, we introduce the model and provide details about the simulation method and the related  protocols. Results are presented and discussed in Section~\ref{sec:results}, while the final section, Section~\ref{sec:conclusion_outlook}, contains the summary of results, concluding remarks, and an outlook to related future investigations.

\section{Model and simulation details}
\label{sec:model_simulations}

For the interactions between the ultrasoft particles we have used the generalized exponential (GEM-$n$) potential, whose functional form is given by (see also \cite{mladek2006formation})

\begin{equation}
    \Phi_{\alpha \beta}(r_{ij}) = \epsilon_{\alpha \beta} \exp[-(r_{ij}/\sigma_{\alpha \beta})^n] .
    \label{gem4-equation} 
\end{equation}
In accordance with previous publications we choose $n = 4$. In Eq. (\ref{gem4-equation}) $r_{ij}$ represents the distance between particles $i$ and $j$ (with $i, j = 1, ..., N$), $N$ being the total number of particles in the system. The indices $\alpha$ and $\beta$ can assume the values A and B, representing the two species of the particles. $\epsilon_{\alpha \beta}$ are the energy parameters and $\sigma_{\alpha \beta}$ are the range parameters of the respective interactions. As we deal with a  (size-)symmetric mixture we impose that $\sigma_{\rm AA} = \sigma_{\rm BB} = \sigma_{\rm AB} \equiv \sigma$. Further, $\epsilon_{\rm AA} = \epsilon_{\rm AB} \equiv  \epsilon$ and $\xi \epsilon = \epsilon_{\rm AB}$ with $\xi = 1.5$. Further $N = N_{\rm A} + N_{\rm B}$, $N_{\rm A}$ and $N_{\rm B} (= N_{\rm A})$ being the number of particles of species A and B, respectively. 

In our simulation-based investigations the potential is truncated at a distance $r_{\rm c}=2.2\sigma$ with $\Phi(r_{\rm c})~\simeq~\epsilon~6.7~\times 10^{-11}$.  All results presented in this contribution are based on simulations of ensembles with $N = 524~288$ particles; we have deliberately chosen such a huge ensemble in an effort to reduce size effects as much as possible. The cubic volume $V$ of the system (with box length $L$) is chosen in such a way that the (reduced, dimensionless) density, defined by $\rho^* = \rho \sigma^3 = \sigma^3 N/V = 2$. Time $t$, temperature $T$, and density $\rho$ are given in units of $\sigma \sqrt{m/\epsilon}$, $k_B T/\epsilon$ and $\rho \sigma^3$, respectively, where $m$ is the mass of the particles and $k_{\rm B}$ is the Boltzmann constant. For simplicity we set henceforward $\epsilon$, $\sigma$, $m$, and $k_{\rm B}$ to unity. 

To simulate the system, (non-equilibrium) molecular dynamics (MD) simulations have been performed in an NVT ensemble, taking advantage of the LAMMPS package \cite{plimpton1995fast}. The temperature of the ensemble is maintained via a dissipative particle dynamics (DPD) thermostat, which is equivalent to the non-conservative portion of a DPD force field \cite{soddemann2003dissipative}. Within DPD, the equations of motion of the particles are given by,

\begin{eqnarray}
  \dot{\bm r_i} &=& \frac{\bm p_i}{m_i} \\
  \dot{\bm p_i} &=& \sum_{j \neq i} \Big [\bm F^{\rm C}_{ij} + \bm F^{\rm D}_{ij} + \bm F^{\rm R}_{ij}], \quad i,j = 1,...,N .
\end{eqnarray}
In the above relations the $\bm r_i$ and $\bm p_i$ are the positions and the momenta of a particle with index $i$. The forces acting on atom $i$ due to other atoms (with index $j$) are given by  the sum over the conservative forces $\bm F^{\rm C}_{ij}$, the dissipative forces $\bm F^{\rm D}_{ij}$, and the random forces $\bm F^R_{ij}$: the $\bm F^{\rm C}_{ij}$ can be calculated from the interparticle potential defined in Eq. (\ref{gem4-equation}). Further, the $\bm F^{\rm D}_{ij}$ and the $\bm F^{\rm R}_{ij}$ are given by,

\begin{eqnarray}
     \bm F^{\rm D}_{ij} &=& - \gamma \omega^{\rm D} (r_{ij}) (\bm{\hat{r}}_{ij} 
      \cdot \bm v_{ij}) \bm{\hat{r}}_{ij} \\
      \bm F^{\rm R}_{ij} &=& \sqrt{2 \gamma k_B T} \omega^{\rm R}(r_{ij}) \Theta_{ij} \bm{\hat{r}}_{ij} .
\end{eqnarray}
$r_{ij} = |\bm r_{ij}| = | \bm r_i - \bm r_j |$ is the distance between particles with indices $i$ and $j$, $\bm{\hat{r}}_{ij} = \bm r_{ij}/|\bm r_{ij}|$ is the unit vector connecting these two particles, and $\bm v_{ij} = (\bm v_i - \bm v_j)$ is the relative velocity between particles $i$ and $j$. Further, $\gamma$ is the friction coefficient, which is set to unity. $\omega^{\rm D}(r)$ and $\omega^{\rm R}(r)$  are distance-dependent weight functions vanishing for $r>R_c$  \cite{espanol1995statistical,hoogerbrugge1992simulating}. The usual choice of the weight functions for the continuous stochastic differential equation within the DPD algorithm is given by,

\begin{equation}
    \omega^{\rm D}(r_{ij}) = [\omega^{\rm R}(r_{ij})]^2 = \begin{cases}
    1 - r_{ij}/R_{\rm c}, & \text{if $0 \leq r_{ij} \leq R_{\rm c}$}.\\
    0, & \text{otherwise}.
  \end{cases}  
\end{equation}
For simplicity we choose the cutoff radii $R_{\rm c}$ for these functions as $R_{\rm c} = r_{\rm c}$. Further, the $\Theta_{ij} = \Theta_{ij}(t)$ represent uniformly distributed random numbers with a Gaussian statistics, i.e., $\langle \Theta_{ij}(t) \rangle = 0$ and $ \langle \Theta_{ij}(t) \Theta_{kl}(t') \rangle = (\delta_{ik}\delta_{jl} + \delta_{il} \delta_{jk}) \delta(t-t')$, with the brackets denoting ensemble averages. 
 
Initially  all the A and B particles are randomly distributed within our simulation box; due to high density of our system it is very likely that the particles overlap. From this initial configuration we start our simulation and equilibrate it during  $10^6$ MD steps $\Delta t$ at a temperature $T=3.0$; for the time increment we have used $\Delta t=0.005$. The equilibrated system is then simulated over $10^6 \Delta t$, storing configurations in intervals of $10^5 \Delta t$. All these configurations serve in the following as independent initial configurations in subsequent MD runs. The system is equilibrated over $10^6 \Delta t$ at the respective temperatures $T$. Eventually the equilibrated systems are simulated over $10^6 \Delta t$ in order to obtain the required information of the system.

\section{Results}
\label{sec:results}

In this Section, we present our results for the system (as specified in the preceding Section \ref{sec:model_simulations}). We will start by characterizing the morphologies of the system that we have obtained across different temperature ranges; we will then proceed to a discussion on the  impact of a quenching process (from high to a low, sub-critical temperature) on the properties of the system.

\subsection{Equilibrium morphologies}
\label{subsec:eq_str}

We first examine the equilibrium morphologies of the system as they occur at different temperatures. In Fig.~\ref{fig:fig1_snapshot}, we show snapshots of the system in different perspectives views (as specified in the caption) at three different temperatures, $T$ = $1.8, 1.4$, and $1.0$. At $T$ = $1.8$ and $1.4$, the two components of the system form a homogeneously mixed phase, while at $T = 1.0$, a clear spatial separation into an A- and a B-rich phase becomes visible.  

\begin{figure}[h!]
\centering
\includegraphics[width=0.45\textwidth]{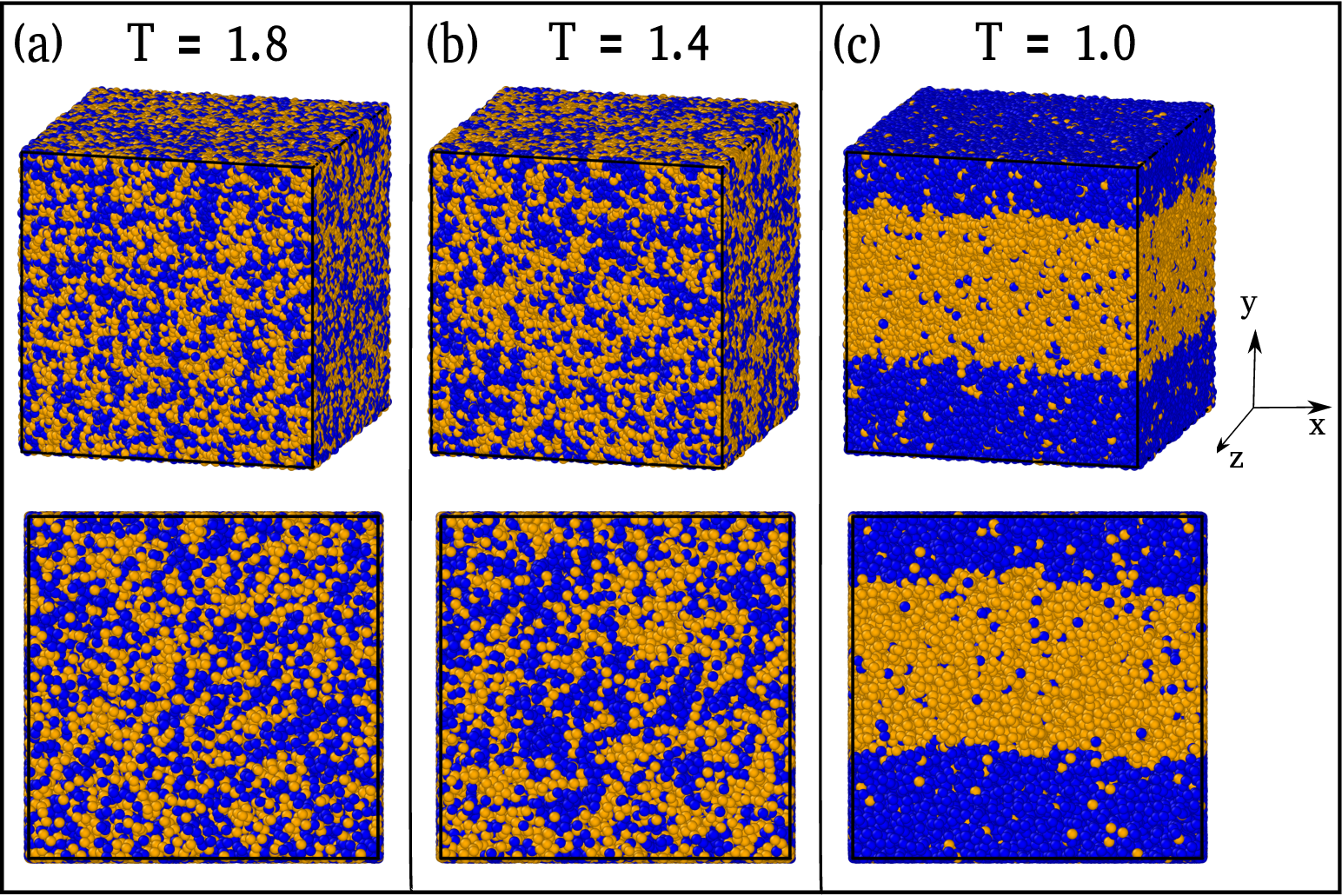}
\caption{Snapshots of the system at hand at different temperatures. Upper panels: perspective views of snapshots of the system at (a) $T = 1.8$, (b) $T = 1.4$, and (c) $T= 1.0$. Lower panels: cross sections of the respective upper panels, taken at $z = L/2$. A- and B-particles are shown in yellow and blue, respectively.}
\label{fig:fig1_snapshot}
\end{figure}
 
In an order to characterize the structure of the system, we have calculated the radial distribution functions $g_{\alpha \beta}(r)$ ($\alpha, \beta$ = A or B) \cite{hansenmcdonald} of the two species, which are defined as,
 
\begin{eqnarray}
\label{eq_gofr}
g_{\alpha \beta}(r) = \frac{N}{\rho N_\alpha N_\beta} \Big< \sum_{i=1}^{N_\alpha} \sum_{j=1}^{N_\beta}{\vphantom{\sum}}' \delta(r-|\bm r_i - \bm r_j|) \Big>   ;
\end{eqnarray}
where the brackets denote an ensemble average and the prime indicates that the contribution is not considered in case $\alpha = \beta$ and $i = j$.

\begin{figure}[h!]
\centering
\includegraphics[width=0.45\textwidth]{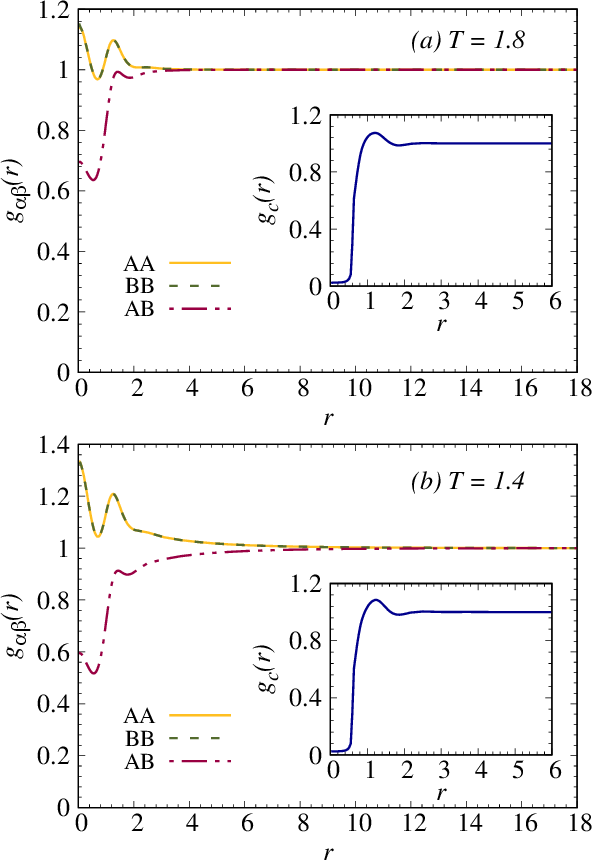}
\caption{Radial distribution functions $g_{\alpha \beta}(r)$ as defined in Eq. (\ref{eq_gofr}) as functions of the distance $r$ for different index combinations (i.e., AA, BB, and AB -- as labeled) of the system at hand, calculated for two different temperatures: panel (a) -- $T = 1.8$ and panel (b) -- $T = 1.4$. Inset: radial distribution functions $g_{\rm c}(r)$ of the centers of mass of the clusters -- see text.}
\label{fig:fig2_gofr}
\end{figure}

The $g_{\alpha \beta}(r)$ are depicted in Fig.~\ref{fig:fig2_gofr} as functions of distance $r$ for two temperatures (as labeled) in the homogeneous, mixed phase (cf. panels (a) and (b) in Fig.~\ref{fig:fig1_snapshot} for the corresponding snapshots). The  $g_{\alpha \beta}(r)$ show the typical behaviour as encountered in a liquid phase: the functions oscillate at small $r$-values and then tend towards unity at large distances. At small distances (i.e., for $r\lesssim 1$), the $g_{\alpha \beta}(r)$ assume finite values (close to unity), indicating the presence of clusters of overlapping particles, a typical feature occurring in systems of ultrasoft particles.

As systems of ultrasoft particles are prone to form clusters at any density \cite{mladek2006formation,mladek2007clustering,likos2007ultrasoft,zhang2010reentrant,mladek2007phase,likos2008cluster,coslovich2011hopping,shrivastav2020,gaurav2020}, we have to properly specify what we mean with a cluster in the disordered phase. To this end we have defined a distance $r_{\rm cl}$, being the position of the first (local) minimum in the $g_{\alpha \beta}(r)$ -- see Fig.~\ref{fig:fig2_gofr} -- as a rough estimate for the spatial extent of a cluster; this distance amounts to $r_{\rm cl} \simeq 0.6$. For all particles pertaining to the same cluster we then define the center of mass, $r_{\rm CM}$, of this clusters of overlapping particles via,

\begin{eqnarray}
r_{\rm CM} = \frac{1}{N_{\rm c}} \sum_{i=1}^{N_{\rm c}} {\bm r_i} .
\end{eqnarray}

In contrast to the partial radial distribution functions $g_{\alpha \beta}(r)$, the radial distribution function of the clusters, $g_{\rm c}(r)$, plotted in the insets of the panels of Fig. \ref{fig:fig2_gofr}, vanishes at small distances, akin to the radial distribution function of a system with steeply repulsive interactions \cite{hansenmcdonald}. This function indicates that the centers of mass of the clusters do not overlap, i.e., that the clusters behave essentially as strongly repulsive, effective particles.

\subsection{Phase separation and critical behavior}
\label{phase_sep}

In an effort to characterize the morphologies of the system (as they have already been shown in Fig.~\ref{fig:fig1_snapshot} for different temperatures) we subdivide the simulation box into cubic cells of length $2\sigma$ and count the number of A-particles ($N_{\rm A}^i$) in the cell with index $i$ and relate this number to the total number of particles contained in this cell, i.e., $(N_{\rm A}^i + N_{\rm B}^i)$; the fraction of these two numbers is the local concentration $\Psi_i$ (which might also be termed $x_{\rm A}^i$) \cite{shaista2007crossover} \footnote{At this point we note that the actual size of the small cells is essentially arbitrary; this size can be fine-tuned -- if required -- in order to obtain good quality data at a coarse-grained level without compromising the results.}
 \begin{eqnarray}
 \label{eq_orp}
 \Psi_i = \frac{N_{\rm A}^{i}}{N_{\rm A}^{i} + N_{\rm B}^{i}} ;
 \end{eqnarray}
Obviously, $\Psi_i \in [0, 1]$.
A typical color-coded map of $\Psi_i$ is shown in Fig.~\ref{fig:density-colormap} for selected snapshots, taken at three different temperatures (as labeled): the upper panels show again perspective views of the simulation box, while the lower panels display cross sections of the respective snapshots, taken at $z = L/2$.

In an effort to learn more about the phase separation of the system into A- and B-rich phases, we have calculated the probability distribution of the $\Psi_i$ in the system at different temperatures, which we term $P(x_{\rm A})$. This is done by subdividing the range $[0,1]$ into 500 bins and by recording the number of occurrences of $\Psi_i$ in each of these pin; in this manner we obtain -- after normalizing -- $P(x_{\rm A})$. The function $P(x_{\rm A})$ is shown in the top panel of Fig.~\ref{fig:pd-coex}, as obtained for different temperatures (as labeled). Obviously, in the homogeneous phase (occurring at high temperatures), $P(x_{\rm A})$ shows a pronounced, symmetric peak centered at $x_{\rm A} = 0.5$. In contrast, in the heterogeneous regime (i.e, at low temperatures), two peaks appear (which are located at $x_{\rm A}$-values symmetric to $x_{\rm A} = 0.5$): here, the peak at the smaller $x_{\rm A}$-value corresponds to the A-poor (or B-rich) phase, while the peak at the larger $x_{\rm A}$-value corresponds to the A-rich (or B-poor) phase. The fact that the distribution function shows two pronounced peaks (and vanishes in between) for the lowest temperature (i.e., at $T = 1.0$) provides evidence that the system is clearly phase separated into regions, hosting either a B-rich (peak at $x_{\rm A} \simeq 0.06$) and an A-rich phase (peak at $x_{\rm A} \simeq 0.94$).

If we then plot the $x_{\rm A}$-values of the peaks in $P(x_{\rm A})$ for the different temperatures we obtain the phase diagram, shown in the bottom panel of Fig.~\ref{fig:pd-coex}. Below the critical temperature, $T_{\rm c}$, we see the characteristic coexistence line, separating the two above mentioned phases.

\begin{figure} [htb]
\centering
 \includegraphics[width=0.45\textwidth]{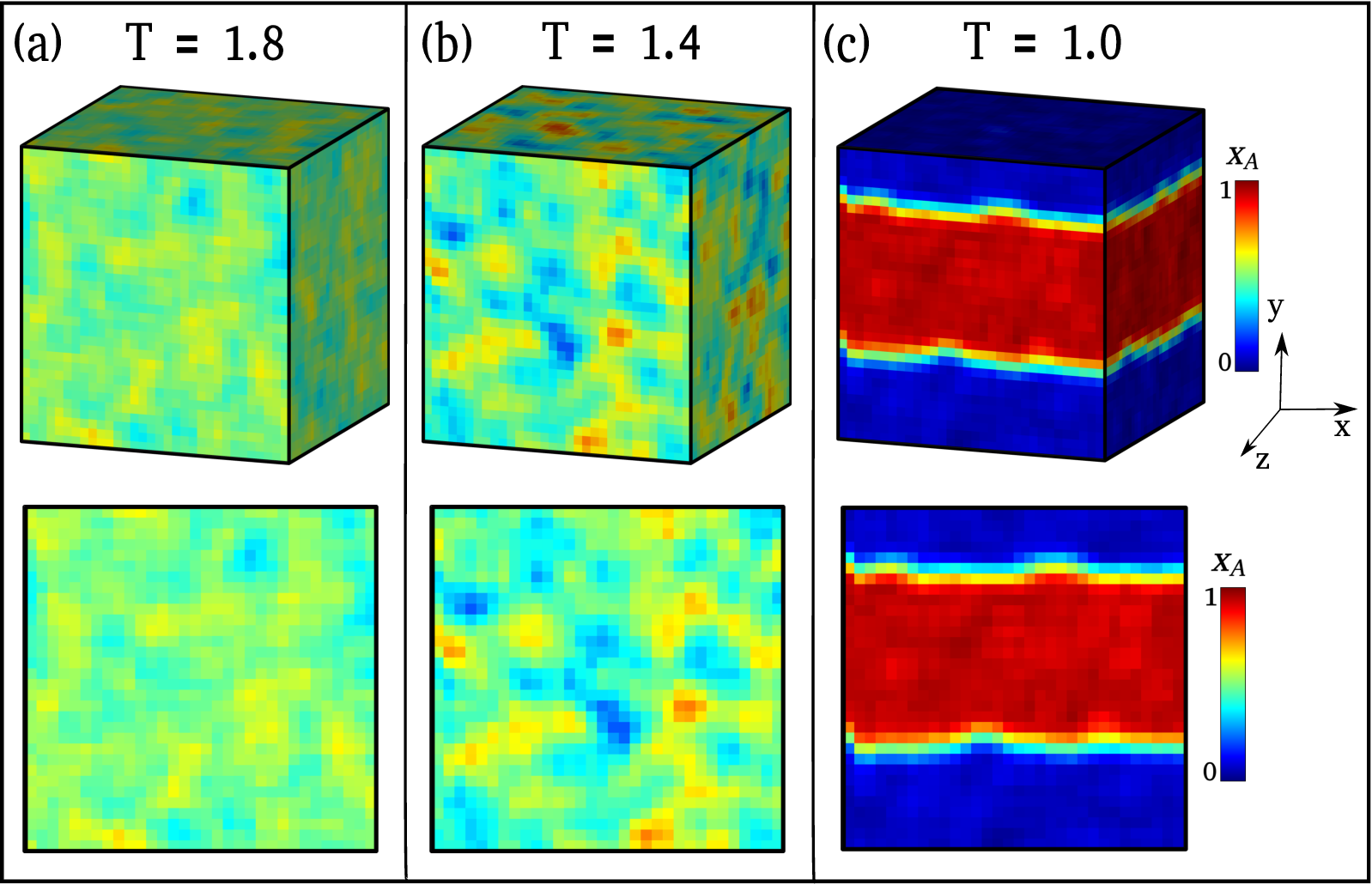}
 \caption{Color-coded representation of $\Psi_i$ (as defined in the text) for selected snapshots. Upper panels: perspective views of snapshots of the system at (a) $T = 1.8$,  (b) $T = 1.4$,  and (c) $T= 1.0$. Lower panels: cross sections of the respective upper panels, taken at $ z = L/2$. The actual value of $\Psi_i \in [0,1]$ of a cell with index $i$ can  be extracted from the respective color-codes, depicted on the right hand side of the snapshots.}
\label{fig:density-colormap}
\end{figure}

The critical temperature $T_{\rm c}$ can be extracted from these data by fitting the difference in the concentrations of the two coexisting phases (indices '(1)' for the B-rich and '(2)' for the A-rich phase, respectively), i.e., $(x_A^{(2)} - x_A^{(1)})$ (as obtained from the positions of the maxima of the distribution function $P(x_{\rm A})$ with the usual functional form \cite{subir2006static}:
\begin{eqnarray}
\label{eq_tc}
x^{(2)}_A - x^{(1)}_A = B(1-T/T_c)^\beta ;
\end{eqnarray}
here, $B$ and $T_{\rm c}$ are fitting parameters, anticipating that the phase separation scenario at hand pertains to the 3D-Ising universality class, i.e., $\beta \simeq 0.325$. The inset of the bottom panel of Fig.~\ref{fig:pd-coex} provides evidence that $(x_A^{(2)} - x_A^{(1)}$ can indeed be nicely fitted via Eq.~(\ref{eq_tc}); the fitted values are:  $B=3.52\pm0.05$ and $T_{c} = 1.351\pm0.001$. 
\begin{figure} [htb] 
\centering
\includegraphics[width=0.45\textwidth]{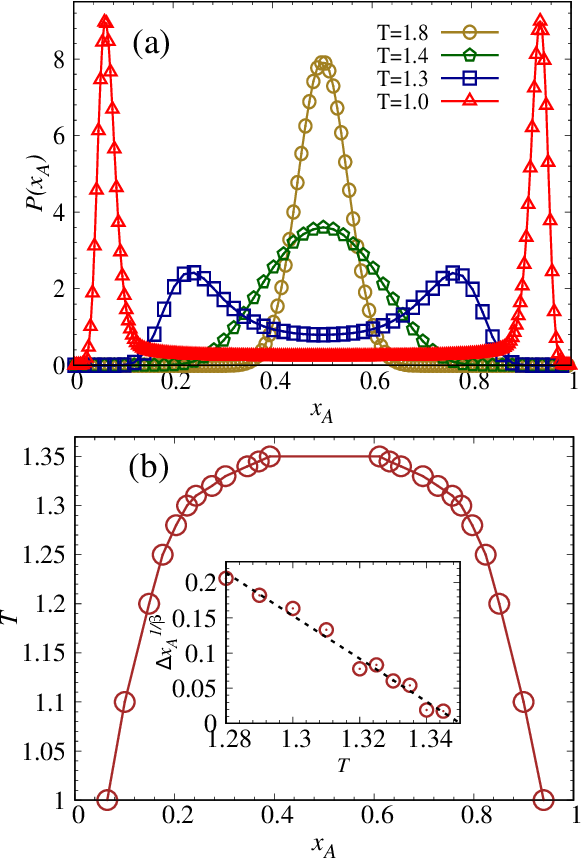}
\caption{Top panel (a): probability distribution, $P (x_{\rm A})$ (as defined in the text) as a function of $x_{\rm A}$ for selected, different temperatures (as labeled). Bottom panel (b): symbols -- values of the coexistence densities $x_{\rm A}^{(1)}$ and $x_{\rm A}^{(2)}$ (as identified as maxima of $P(x_{\rm A})$) for the different (sub-critical) temperatures investigated in this study. The red line displays the function specified in Eq. (\ref{eq_tc}), using the fitting parameters given in the text. The inset shows $\left[ 
x_A^{(2)} - x_A^{(1)} \right]^{(1/\beta)}$ as a function of temperature $T$; the dotted line displays again the function specified in Eq. (\ref{eq_tc}).}
\label{fig:pd-coex}
\end{figure}

\subsection{Phase separation dynamics}

We proceed by exploring the non-equilibrium dynamics of the mixture when quenched from a high temperature to a low, subcritical temperature. We start off from the equilibrated mixture at $T=3.0$ which is in a homogeneous phase. We instantaneously quench the system down to $T=0.68 ~ (\simeq 0.5T_{c})$, i.e., to a state located inside the coexistence region of the phase diagram. The two components of the system demix, forming A-rich and B-rich domains. In an effort to visualize the emerging morphologies, we assign a value $+1$ to A-rich (sub-)cells (introduced above) and a value $-1$ to B-rich (sub-)cells. Snapshots of the mixture and their coarse-grained versions (with binary values) at $T = 0.5~T_{\rm c}$ are shown in Figs.~\ref{fig:quench_snapshot} and \ref{fig:quench_snapshot_coarse}, respectively. In each of the figures panels (a), (b), and (c) show the snapshot of the mixture in after different time-spans after the quench (as labeled). The respective top panels show the the snapshots in perspective views, while the bottom panels are cross sections of the respective upper panels, taken at $z = L/2$.

\begin{figure}[h!]
\centering
\includegraphics[width=0.45\textwidth]{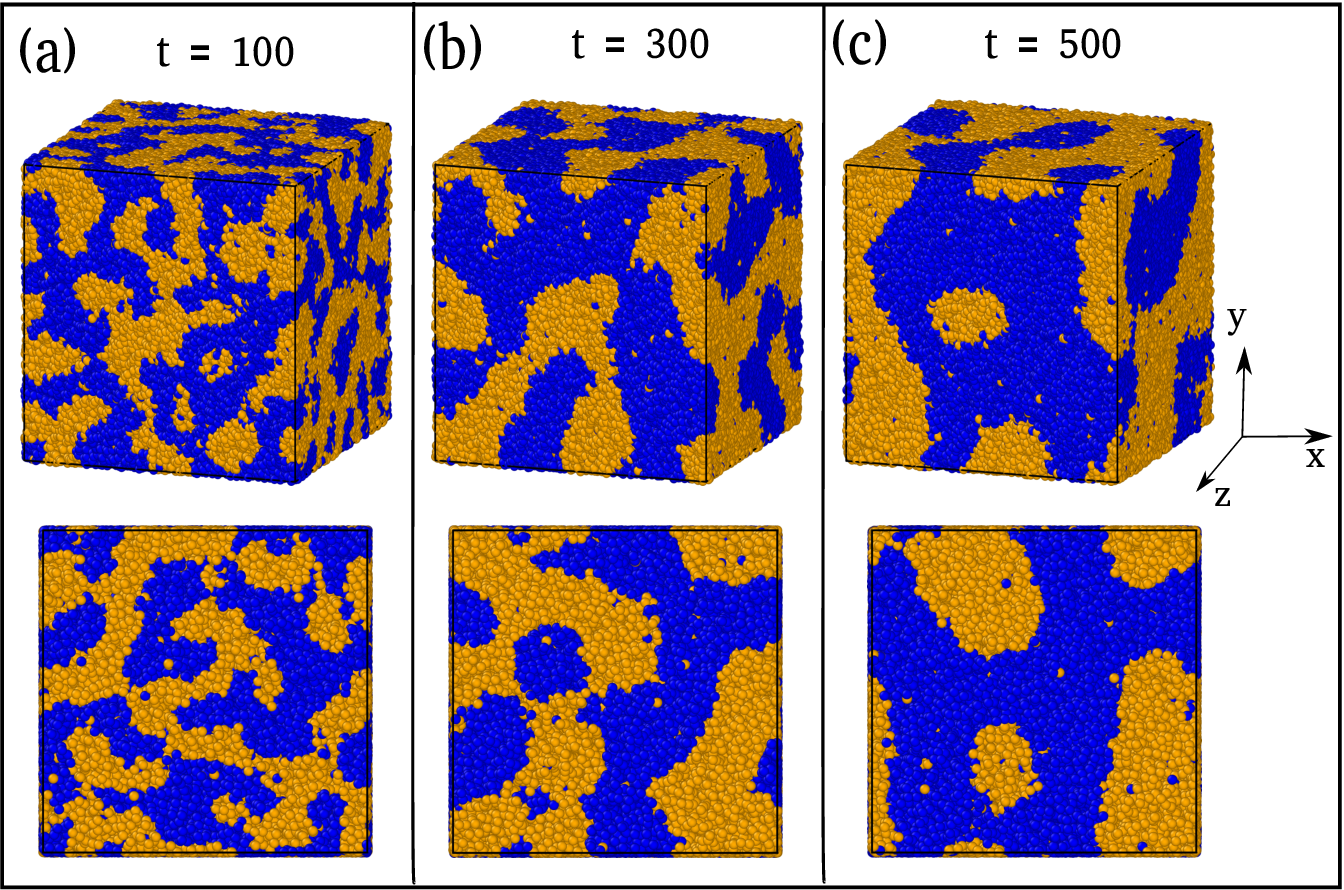}
\caption{Snapshots of the system at different times $t$ after the system has been quenched down to $T = 0.5~T_{\rm c}$ (as detailed in the manuscript). Upper panels: perspective views of snapshots of the system at (a) $t = 100$, (b) $t = 300$, and (c) $t= 500$. Lower panels: cross sections of the respective upper panels, taken at $z = L/2$. A- and B-particles are shown in yellow and blue, respectively.}
\label{fig:quench_snapshot}
\end{figure}

\begin{figure}[htb]
\centering
\includegraphics[width=0.45\textwidth]{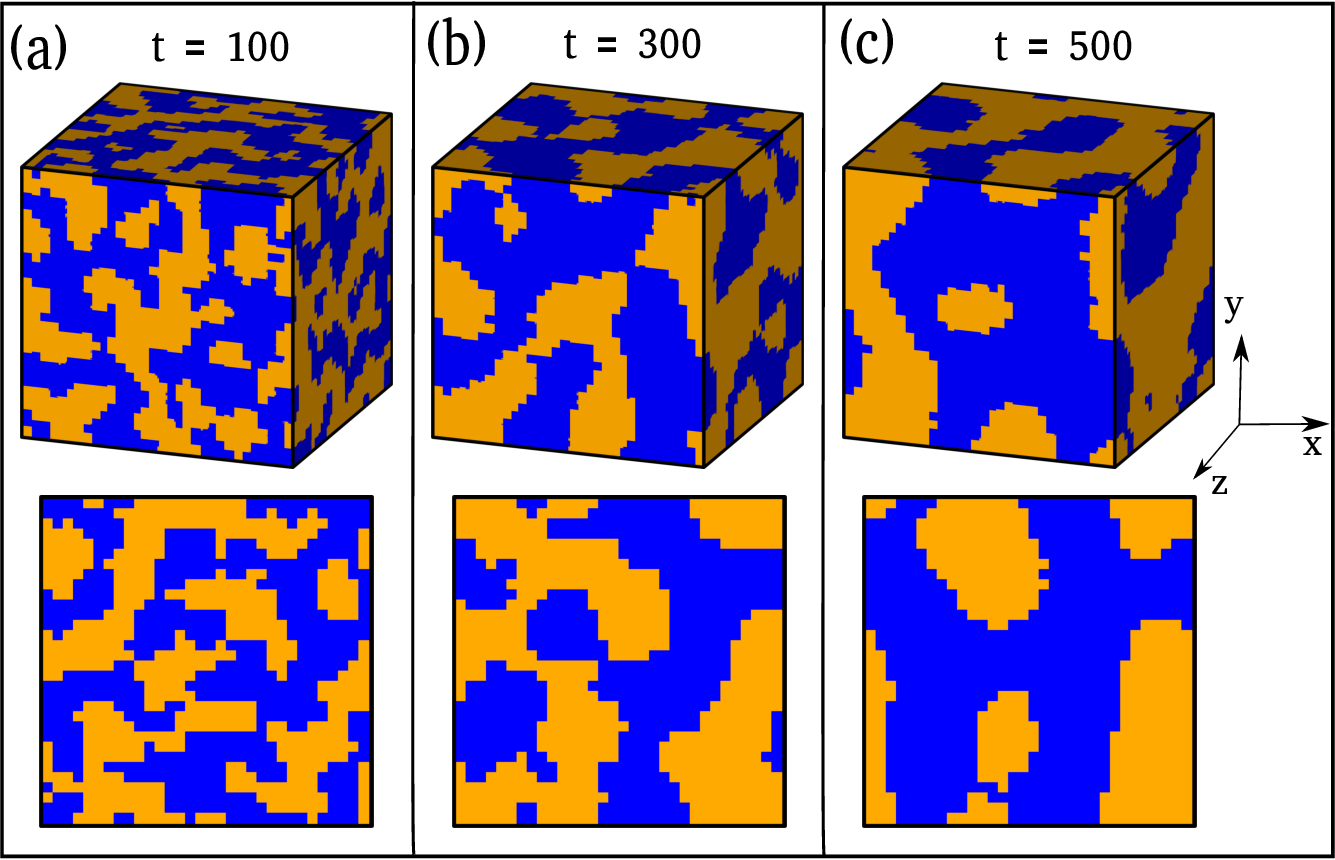}
\caption{Same as Fig.~\ref{fig:quench_snapshot}, now in a coarse-grained representation, based on the cubic (sub-)cells (see text).}
\label{fig:quench_snapshot_coarse}
\end{figure}

\begin{figure} [htb]
\centering
\includegraphics[width=0.45\textwidth]{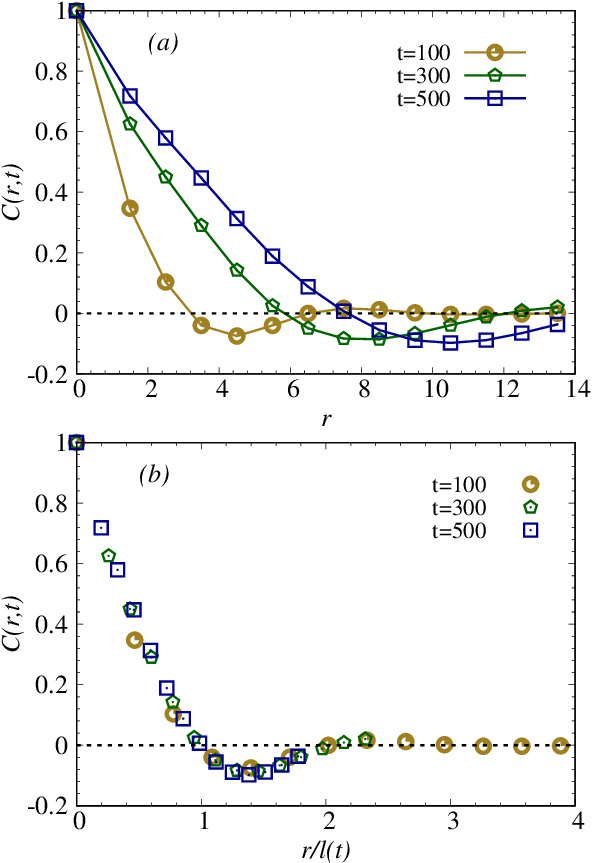}
\caption{Top panel: correlation function $C({\rm r}, t)$ as defined in Eq. (\ref{eq_corrfn}) as a function of $r$, evaluated for different values of time $t$ (as labeled) after the quench; the considered temperature is$T = 0.5 T_{\rm c}$. Bottom panel: correlation function $C({\rm r}, t)$, shown in the top panel, now as a function of the scaled distance $r/l(t)$, as defined in the text.}
\label{fig:correlation_function}
\end{figure}

We track the spatial evolution of these domains with time $t$ by calculating the correlation function $C(\bm r, t)$ of the parameter $\Psi({\bm r}, t)$ (previously introduced as $\Psi_i$) -- ${\bm r}$ being the position vector of the sub-cell within the simulation box --, evaluated at the time $t$; where the time-scale starts once the quench of the system has been realized; the above function is defined as:

\begin{eqnarray}
\label{eq_corrfn}
\begin{split}
C(\bm r, t) &= \frac{1}{V} \int_V d\bm R \Big [ \Big< \psi(\bm R,t) \psi(\bm R+\bm r,t) \Big>  \\ 
& - \Big< \psi(\bm R,t)\Big> \Big<\psi(\bm R+\bm r,t) \Big>\Big] 
\end{split}
\end{eqnarray}
This function correlates the quantity $\Psi (\bm r, t)$, evaluated for two different sub-cells within the simulation box at the same time $t$: $\bm R$ and $(\bm R + r)$ are the positions of two sub-cells and $\bm r$ is the vector between these two cells points. 

The above correlation function $C(\bm r, t) = C(r, t)$, calculated for different values of $t$ (namely $t=100$, $300$ and $500$) is shown in the top panel of Fig.~\ref{fig:correlation_function} at a temperature $T=0.5~T_c$. $C(r, t)$ is displayed as a function of $r$ (with the range in $r$ being limited by the simulation box). The correlation function initially decays linearly for all $t$-values considered, then fluctuates around zero, and eventually tends towards zero.

The correlation functions collapse -- when their $r$-dependence is appropriately scaled via a time-dependent scaling factor $l(t)$ -- on a single master curve, as shown in the bottom panel of Fig.~\ref{fig:correlation_function}. Thus, 

\begin{equation}
\label{eq_corrfn_1}
C(r, t) = g\Big(\frac{r}{l(t)}\Big) ,
\end{equation}
introducing thereby the function $g(r)$. The spatial scaling factor for the distance is the (time-dependent) average size of the domains, $l(t)$, which can be estimated as follows: a measure for $l(t)$ is obtained by identifying the first zero of the function $C(r, t)$ (see, e.g., \cite{shaista2007crossover}). This length is shown in Fig.~\ref{fig:l_t} as a function of $t$.

\begin{figure}[h!]
\centering
\includegraphics[width=0.45\textwidth]{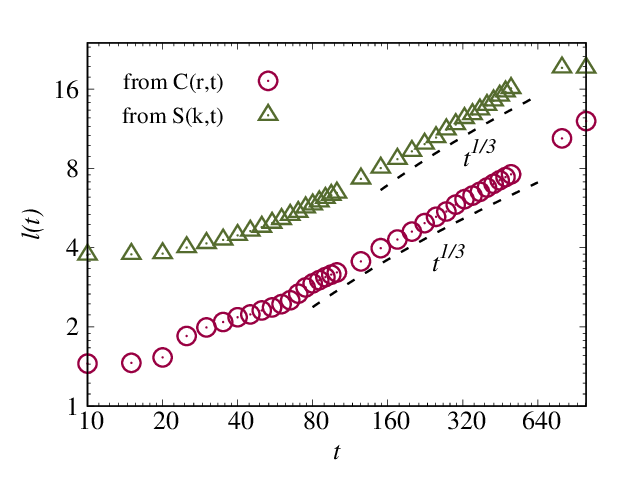}
\caption{$l(t)$ as a function of $t$ in a double-logarithmic representation, calculated at the temperature $T=0.5T_{\rm c}$. Data are obtained -- as labeled -- from the correlation function $C(r, t)$ -- as defined in Equation (\ref{eq_corrfn}) -- and via the time-dependent structure factor -- as defined in Equation (\ref{structure_factor}) -- as labeled by the different symbols. For details see text. The broken lines indicate a $t^{1/3}$ behaviour of $l(t)$, as predicted by Lifshitz and Syolov (see text).}
\label{fig:l_t}
\end{figure}


For $t \gtrsim 80$ the length $l(t)$ shows a power-law behaviour with an exponent $1/3$ \cite{hohenberg1977theory, bray1994review}, a behaviour which is consistent with the theoretical predictions for conserved systems, found, e.g., by Lifshitz and Slyozov \cite{LS1961kinetics}.

We further define the spatial Fourier transform of the correlation function $C({\bm r}, t]$ via

\begin{eqnarray}
\label{eq_strfac}
S(\bm k,t) = \int d \bm r e^{i \bm k \cdot \bm r} C(\bm r,t) = l(t)^d {\tilde g}(k l(t))  .
\label{structure_factor}
\end{eqnarray}
$S(\bm k, t) = S(k, t)$ can be considered as a time-dependent structure factor. $d (= 3)$ represents the dimension of the system Further, $\bm k$ is the wave vector and $\tilde g(p)$ is the Fourier transform of $g(x)$, i.e.,

\begin{eqnarray}
{\tilde g}(p) = \int d \bm x e^{i \bm p \cdot \bm x} g(x)
\end{eqnarray}

The panels of Fig.~\ref{fig:fig_str} show the function $S(k, t)$, either as a function of $k$ -- top panel -- or of the scaled wave vector, i.e. $k l(t)$ -- bottom panel; $S(k, t)$ is shown for three different values of time $t$ (as labeled), evaluated at the temperature $T = 0.5T_{\rm c}$; note the semi-logarithmic representations. The unscaled functions show pronounced peaks at $k$-values, which indicate (via their respective inverse values) the size of the above mentioned domains. Replotting $S(k, t)$ as a function of the scaled wave vector, i.e., of $k l(t)$, we find that the functions, calculated for different $t$-values collapse on a single master curve -- see bottom panel of Fig.~\ref{fig:fig_str}, showing thus essentially the function $\tilde g(k l(t))$ as defined above in Equation (\ref{structure_factor}). Further we observe that the $S(k l(t), t)$ show for $k l(t) \gtrsim 1.6$ a power-law type decay with an exponent $4 (= d +1)$. This observation is consistent with Porord's law \cite{porod1982}, which predicts for larger $k$-values a decay of the scaled function $S(k l(t), t)$ with an exponent $(d+1)$ signifying the presence of sharp interfaces in the mixture.

\begin{figure}[h!]
\centering
\includegraphics[width=0.45\textwidth]{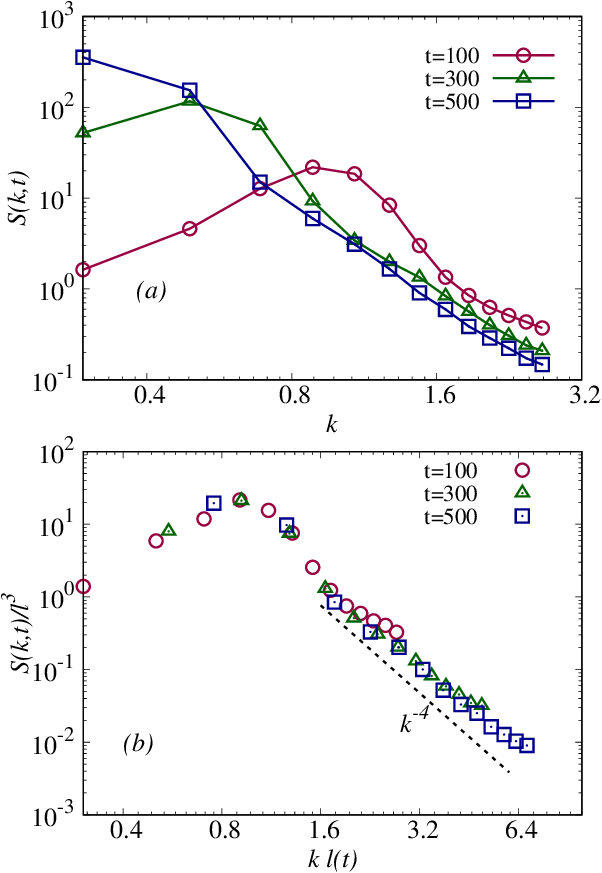}
\caption{Time-dependent structure factor $S(k, t)$ as defined in Eq. (\ref{structure_factor}) as a function of the wave vector $k$ -- top panel -- and of the scaled wave vector $k l(t)$ -- bottom panel; $S(k, t)$ is evaluated for different values of $t$ (as labeled). The dotted line in the bottom panel indicates a power-law decay in $k$ with $\sim k^{-4}$. Note the semi-logarithmic presentation of the data.}
\label{fig:fig_str}
\end{figure}




\section{Conclusions and Outlook}
\label{sec:conclusion_outlook}

With the help of extensive molecular dynamics simulations (both with respect to ensemble size and simulation length), carried out in an NVT ensemble we have studied the liquid-liquid phase separation scenario in an equimolar, size-symmetric binary mixture of ultrasoft particles (with species labeled A and B). Consistent with previous publications we use the generalized exponential model of index $n = 4$ for the interparticle interaction.  In an effort to keep the temperature fixed we use a DPD thermostat, inbuilt in the LAMMPS program package. While the interactions among like particles are identical, the cross interactions is by a factor of 1.5 stronger, i.e., more repulsive, inducing thereby the desired phase separation scenario.

In a first step we have investigated the equilibrium morphologies of the system via static correlation functions and trace out the phase diagram of the system via calculating and analysing the distribution functions of the local concentration of particle species~A. An analysis of these distribution functions in combination with an extrapolation of the coexistence densities (and anticipating that the phase separation scenario pertains to the Ising 3D universality class) yields a critical point at a temperature $T_{\rm c} = 1.35$. Snapshots of the system confirm the expected behaviour at super-critical, at close-to-critical and at sub-critical temperatures. The structure of the system is analysed for a wide range of temperatures (both above and below $T_{\rm c}$) in terms of the (partial) radial distributions and the radial distribution function of the centers of mass of the clusters of overlapping particles formed in the system. At supercritical temperatures the radial distribution function of the clusters shows the expected behaviour: a finite value at short distances, indicating the formation of clusters of mutually overlapping particles as well as  oscillations at long distances. The cluster correlation function shows a behaviour akin to a system of strongly repulsive particles, indicating that the clusters behave as effective, mutually repulsive particles.

In a further step we have explored the dynamics of the phase separation process within the system. To this end we have rapidly  (i.e., instantaneously) quenched the ensemble from a high temperature to the subcritical temperature $T = 0.5 T_{\rm c}$. In a time-dependent process, the initially homogeneously mixed system then phase separates into A- and B-rich phases, a process which is tracked over time via the time- and space-dependent correlation function $C(\rm r, t)$ of the local concentration: to this end we have subdivided the simulation box in small cubic sub-cells and have recorded the local concentration of either species. This function correlates at the same instant the concentrations in two sub-cells, separated by a distance ${\rm r}$. As the properties of the system is tracked along the time scale, a time-dependent (average) size of clusters of particles, $l(t)$, which grows (from intermediate times onwards) in time via a power law (consistent with the predictions of Lifshitz and Slyozov). Scaling the distance between two sub-cells via $l(t)$ makes the correlation functions calculated for different time instants collapse on a single, time-independent master curve. Likewise, the time-Fourier transform of $C(\rm r, t)$, the time-dependent structure factors collapse -- when the wave-vector is scaled by the cluster size -- on a single master curve which shows at large wave-vectors a power-law decay, as predicted by Porod's law. 

With the knowledge achieved from this particular, symmetric case we are ready to push forward our investigations on binary mixtures of ultrasoft particles in two directions: on one side we plan to introduce asymmetry into the system, namely both with respect to the size of the particles as well as to the interaction strength. On the other side we will expose the system to other non-equilibrium conditions, such as the exposure to shear forces. Investigations in both directions are currently on their way.

\newpage








\bibliography{sorsamp}

\end{document}